\documentclass[twocolumn,pre,aps,showpacs]{revtex4}
\usepackage{graphicx}
\usepackage{amssymb}
\usepackage{amsmath}
\usepackage{hyperref}
\usepackage{latexsym}

\def\be{\begin{equation}}
\def\ee{\end{equation}}
\def\bea{\begin{eqnarray}}
\def\eea{\end{eqnarray}}

\draft

\begin{document}
\title{Mechanical stability of bipolar spindle assembly}
\author{Paolo Malgaretti$^{1,2}$ and Sudipto Muhuri$^{3}$}
\affiliation{$^{1}$ Max-Planck-Institut f\"{u}r Intelligente Systeme, D-70569,Stuttgart, Germany\\
$^{2}$ Institut f\"ur Theoretische Physik, Universit\"{a}t Stuttgart, D-70569, Stuttgart, Germany\\
$^{3}$ Department of Physics, Savitribai Phule Pune University, Ganeshkhind, Pune 411007, India}

\begin{abstract}
Assembly and stability of mitotic spindle  is governed by the  interplay of  various intra-cellular forces, e.g. the forces generated by motor proteins by sliding overlapping anti-parallel microtubules (MTs) polymerized from the opposite centrosomes, the interaction of kinetochores with MTs, and the interaction of MTs with the chromosomes arms. We study the mechanical behavior and stability of spindle assembly within  the framework of a minimal model which includes all these effects. For this model, we derive a closed--form  analytical expression for the force acting between the centrosomes as a  function of their separation distance and we show that an effective potential can be associated with the interactions at play. We obtain the stability diagram of spindle formation in terms of parameters characterizing the strength of motor sliding, repulsive forces generated by polymerizing MTs, and the forces arising out of interaction of MTs with kinetochores. The stability diagram helps in quantifying the relative effects of the different interactions and elucidates the role of motor proteins in formation and inhibition of spindle structures during mitotic cell division. We also predict a regime of {\it bistability}  for certain parameter range, wherein the spindle structure can be stable for two different finite separation distances between centrosomes. This occurrence of bistability also suggests mechanical versatility of such self-assembled spindle structures.
\end{abstract}


\maketitle
The assembly of the mitotic spindle is a key event in cellular division. During mitotic cell division, the two centrosomes within the cell serves as the poles and  nucleating sites for microtubules (MTs).  The polymerizing MTs from the opposite centrosomes overlap, leading to the formation of spindle structure during metaphase \cite{cell,howard,raja,ferenz} (see Fig.~1). These polymerizing centrosomic MTs interact with the chromosome arms and generate an effective repulsive force between the centrosomes - also called {\it polar ejection forces}~\cite{ A1,A2}. Some of these MTs also interact with  specialized cellular structures of kinetochores forming kinetochore microtubules generating an additional tension between the two centrosomes~\cite{C1,C2,C3}. Finally, MTs from the opposite centrosomes overlap in the spindle mid-zone. These anti-parallel overlapping MTs are crosslinked by motor proteins such as Eg5 and dynein which attach to the two overlapping MTs and exerts forces which tends to slide the anti-parallel MTs with respect to each other. 
\begin{figure}[t]
\centering
\includegraphics[scale=0.3]{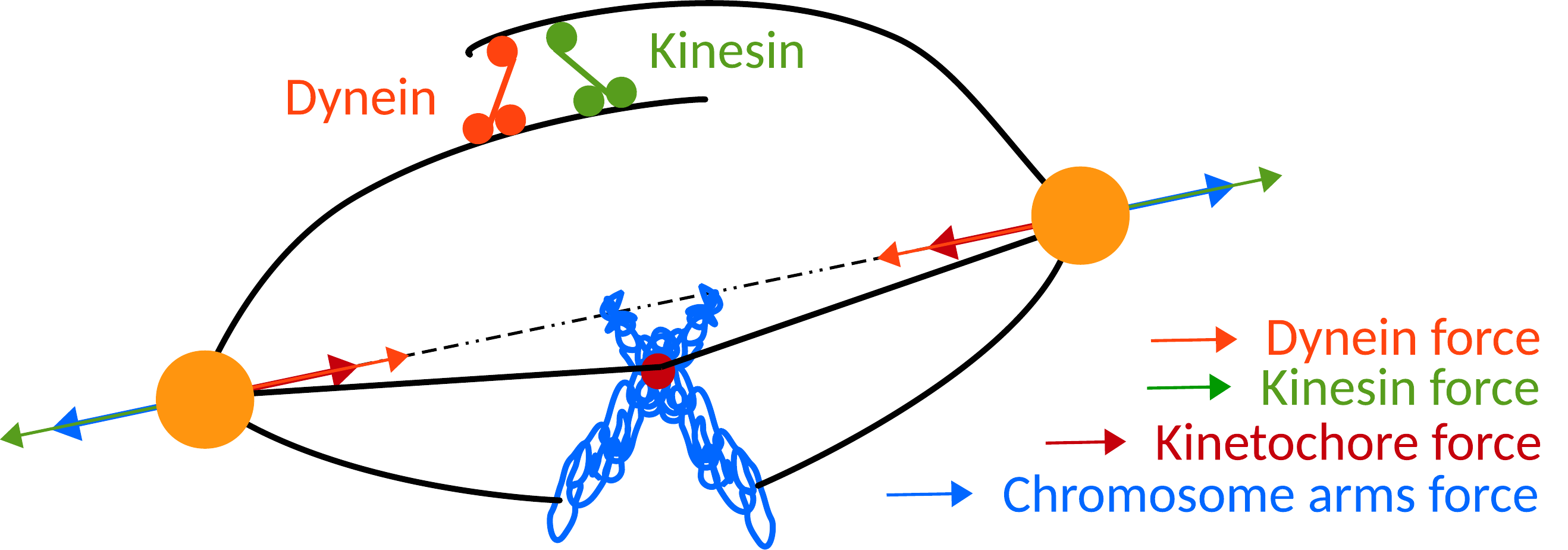}
\caption{Schematic representation of the forces acting along the axis joining the centrosomes: the active displacement of molecular motors leads to net inward or outward sliding forces. The pressure of polymerizing MTs on chromosomes arms leads to an effective repulsive force, while interaction of MT with kinetochores leads to an effective attraction between the centrosomes. } 
\label{fig-1}
\end{figure}
While kinesin motors exert a force on the overlapping MT filaments tending to increase the centrosome separation distance, dynein motors exert a force which tend to decrease the separation between centrosomes ~\cite{B1,B2,B3,raja}.  The net force acting on spindle assembly is the combined effect of all these different forces at play. Apart from these interactions, some of the MTs also interact with the cell cortex~\cite{astral-mt}.  Typically when a stable bipolar spindle configuration is attained as a result of these interactions, chromosomes are localized and aligned on a plane whose normal is aligned along the axis of the spindle, and which lies about the midpoint between centrosomes. Even though each of these interaction forces are  itself the result of complex {\it active}  phenomena, Ref.~\cite{raja,ferenz} have proposed a simple coarse grained model which  takes into account the combined effect of the interaction between the two centrosome and the interactions of the individual centrosomes with the chromosomes. 
 Numerical simulations of this model have shown that the relative strength of the different interactions outlined earlier can regulate the stability of the spindle structure, modulate the stable spindle length and determine the organization of the chromosomes within the spindle structure~\cite{raja,ferenz}. 

In this letter, we use the minimal  model of Ref.~\cite{raja, ferenz} by assuming that {\it a priori} chromosomes are homogeneously distributed on a disk or a ring lying in a plane perpendicular to the axis of the spindle at the midpoint of the spindle axis. Such a configuration is typical in both {\it in vivo} and {\it in vitro} experiments~\cite{cell} and it has been confirmed also in the numerical studies in Ref.~\cite{raja}. This assumption allows us to obtain a closed-form  analytical expression for the force acting between the centrosomes as a function of their separation distance. 
Mechanical equilibrium of the spindle system requires the overall force to vanish. Further the requirement for  {\it stable} mechanical equilibrium demands that any slight deviation from the equilibrium position should result in restoring force which tends to bring back the system to its original state of mechanical equilibrium. Using the expression for the force and the criterion for mechanical stability, we construct the entire stability diagram of spindle formation which is expressed in terms of  parameters which characterizes the strength of motor sliding, repulsive forces generated by polymerizing MTs, and the tensions exerted by kinetochores. 
The stability diagram helps in quantifying the relative effects of the different processes on the stability of the spindle assembly. 
We find that for sufficiently large attractive sliding forces exerted by crosslinking motors on MTs, stability of the spindle structure is lost. Further, we find that for sufficiently small kinetochore-MT interaction strength,  both bipolar spindle structure and monopolar structure (with zero separation distance between centrosomes) are stable as has been observed in numerical studies of Ref.~\cite{raja, ferenz}. We also find a regime of {\it bistability} of spindle assembly for certain parameter range wherein the spindle can be stable for two different finite separation distance between centrosomes. Moreover, our closed--form expression allows us to derive an effective potential which account for the various interactions at play.  Accordingly, the stability diagram and specifically the occurrence of bistability can be rationalized in terms of an multiple-minima in the  effective potential energy function landscape. Finally  we find that the regime of bistable behavior is accessible for such parameter values  which corresponds to a net attractive  sliding force due to crosslinked motors and thus necessarily implies presence of dynein motors which are capable of generating attractive forces between the centrosomes. 

 We first describe the theoretical framework of the proposed model for spindle formation for the case of a single pair of chromosome. Next we extend this model for the case of multiple chromosomes homogeneously distributed on a disc or a ring in the mid-plane between the two centrosomes and we derive the corresponding stability diagram for the spindle structure. Finally we conclude with discussion on the possible implications of this study in analyzing the mechanical stability of spindles.  

\section{Single chromosome} 
We analyze the stable configurations of the two centrosomes when a single chromosome located at the mid-plane between two centrosomes located at $-x$ and $x$.  The net force due to the interactions of the MTs with the chromosome is directly proportional of the number of MTs reaching the chromosome arms. Typically, the number of MTs reaching the chromosome arms at a distance $x$ is proportional to $e^{-x/L}$~\cite{ferenz, exp-length} where $L$ is mean length of MTs. Accordingly we can write down the expression for the polar ejection force, $F_{pe} = Ae^{-x/L}$, where $A$ is the maximum polar ejection force. The sliding force due to the motors is proportional to the overlapping MTs from the opposite centrosomes~\cite{raja, ferenz}. Thus for centrosomes separated by a distance $2x$, the force due to the motor sliding is $F_{m} = 2Bxe^{-2x/L}$, where $B$ is the net force per unit overlap length of the MTs. The sign of B is positive for net outward forces (when the force due to kinesin motors exceeds the forces of dynein) and it is negative for net inward (attractive) force between centrosomes. The force due to sub-cellular machinery of kinetochore with MTs is attractive~\cite{f-chromo1, f-chromo2}. While in general the attractive force due to the interaction between the kinetochore sub-cellular machinery and MTs depends on the distance of the kinetochore from the centrosome~\cite{raja2}, for the sake of simplicity of the analysis here it is assumed to be constant~\cite{ferenz}. Finally, the role of the interaction of the MTs with the cell cortex is ignored as this interactions does not significantly impact the steady state pattern of the spindles~\cite{raja}. Adding all the different contributions, the expression for the net force between the centrosomes as a function of their separation distance reads~\cite{ferenz}
\begin{equation}
F(x) = Ae^{-x/L} + 2Bxe^{-2x/L} - C 
\label{eq:F1}
\end{equation} 
We use Eqn.(\ref{eq:F1}) to determine the mechanically stable steady states. For this we first note that the condition for mechanical equilibrium implies that for a separation distance $2x = 2x_p$, $F(x_{p}) = 0$. Further for the stable mechanical equilibrium, $\frac{d F}{dx}\big|_{x_{p}} < 0$, which simply means that on slight deviation from equilibrium separation, the restoring forces are such that the system regains its equilibrium configuration. Thus the conditions; $F(x_{p}) = 0$ and $\frac{d F}{dx}\big|_{x_{p}} = 0$ determines the phase boundary, separating linearly stable and unstable regions. In order to simplify the analysis we express  Eqn.~(\ref{eq:F1}) in terms of dimensionless scaled variables $x_{o} = 2x/L$, $F_{o} = F/A$, $C_{o} = C/A$, $B_{o} = \frac{BL}{A}$.  Therefore on the phase boundary,
\small
\begin{eqnarray}
 F_{o}(x_{o})=e^{-x_{p}/2} + B_{o}x_{o} e^{-x_{o}}  - C_{o}=0 \label{eq:scaled-dF1-1}\\
 \frac{dF_{o}}{d x_{o}} =-\frac{1}{2}e^{-x_{o}/2} - B_{o}x_{o} e^{-x_p}  + B_{o}e^{-x_o}=0.\label{eq:scaled-dF1-2}
\end{eqnarray}
\normalsize
\begin{figure}
\includegraphics[height = 2.8in, angle=-90]{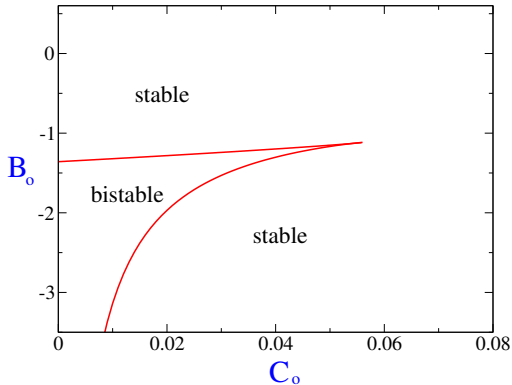}
\caption{Stability diagram for single chromosome expressed in terms of scaled variables. For stable region corresponds to one stable solution of finite separation distance between the two centrosomes. In the bistable region, two different finite separation distances between the the two centrosomes are stable.}
\label{fig:2}
\end{figure}
Using Eq.~(\ref{eq:scaled-dF1-1}) and Eq.(\ref{eq:scaled-dF1-2}), we obtain the expression for the equilibrium separation between the centrosomes at phase boundary as, 
\begin{equation}
x_{o} = 2 \ln \left( \frac{-4B_o}{1 +\sqrt{1 + 16B_{o}C_{o}}}\right)
 \label{eq:scaled-x1}
\end{equation}
Substituting Eq.~(\ref{eq:scaled-x1}) into  Eqs.~(\ref{eq:scaled-dF1-1}),(\ref{eq:scaled-dF1-2}), we obtain the equation for the phase boundary in terms of the scaled parameters $B_{o}$ and $C_{o}$
\begin{eqnarray}
&C_{o}& = - \frac{1 +\sqrt{1 + 16B_{o}C_{o}}}{4B_o}\\
 &+& 2 B_{o} {\left(  \frac{1 +\sqrt{1 + 16B_{o}C_{o}}}{4B_o}\right)}^{2} \ln \left(\frac{-4B_o}{1 +\sqrt{1 + 16B_{o}C_{o}}}\right) \nonumber
\end{eqnarray}
In Fig.~\ref{fig:2} we plot the stability diagram in terms of the scaled variables $B_o$ and $C_o$. As shown in the figure  the bipolar spindle solution is always stable. Moreover, for a range of negative values of $B_o$ which corresponds to motors exerting a net inward force between the centrosomes, and when $C_o$ is  less than a critical value, there is a region of bistability, where two different solutions of spindles, corresponding to two different finite separation distance between centrosomes are stable.

\section{Multiple chromosomes distributed on a disc} 
When many chromosomes are present, the force acting along the line joining the two centrosomes Eq.~(\ref{eq:F1}) can be generalized to:
\begin{equation}
 F= 2Bxe^{-2x/L}+\sum_{i=1}^{N}\frac{x}{R_{i}}\left(Ae^{-R_{i}/L} - C\right)
 \label{eq:my-form0}
\end{equation}
where  $R_i=\sqrt{{x}^{2} + r_{i}^{2}}$ is the distance between the centrosome and the $i$th chromosome lying on disc in the mid-plane between the two centrosomes and $r_{i}$ is the distance of the chromosome from the center of the disc. 
Assuming that $N$ chromosomes are are uniformly distributed on the disc of radius $R_d$, the expression in terms of summation can be converted into an integral expression for the force which reads 
\begin{eqnarray}
F &=&\frac{N x}{\pi R_{d}^{2}}\int_{0}^{R_{d}}\int_{0}^{2\pi}\frac{rdrd\theta}{\sqrt{x^{2}+r^{2}}} \left( A e^{-\sqrt{x^{2}+r^{2}}/L} - C \right)\nonumber\\ &+&2Bxe^{-2x/L} 
\label{eq:my-form}
\end{eqnarray}
which upon integration yields,
\begin{eqnarray}
F&=& 2x\left[Be^{-2x/L}+ \frac{NLA}{R_{d}^{2}}\left( e^{-x/L}-e^{-\sqrt{x^{2}+R_{d}^{2}}/L}\right)\right.\nonumber\\
&-& \left.\frac{NLC}{R_{d}^{2}}\left(\sqrt{x^{2}+R_{d}^{2}}-x\right)\right]
\label{eq:last}
\end{eqnarray}
The net force acting perpendicular to the axis joining the two centrosomes add up to zero due to the radially symmetric arrangement of chromosomes in disc configuration.
The radius of the disk $R_{d}$ is set by the number of chromosomes, $N$, in the disc and by the chromosomal packing fraction that ultimately depends on the mutual interaction between chromosomes. In order to estimate $R_{d}$, we make a crude approximation that each chromosome occupy a surface area $\pi r^{2}_{ch}$ within the disc, where $ r_{ch}= 1 \mu m$\cite{raja} is the radius of a single chromosome. Therefore, $R_{d}\sim r_{ch}\sqrt{N}$.    
\begin{figure}[t]
\centering
\includegraphics[scale=0.27,angle=-90]{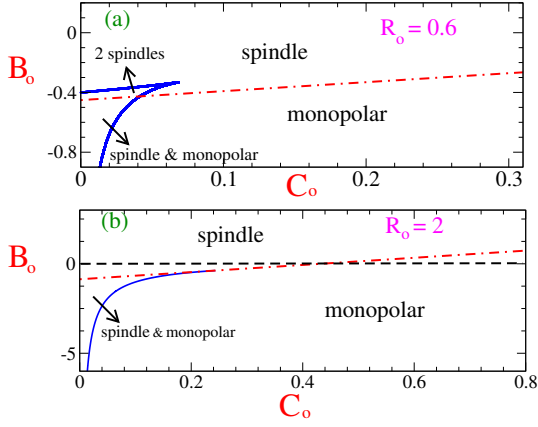}
\begin{center}
\caption{Stability Diagram when the chromosomes are uniformly distributed on a disc of radius $R_{d}$. Here for (a) $R_{o} \equiv R_{d}/L = 0.6$, and for (b) $R_{o} \equiv R_{d}/L = 2$.} 
\label{fig-3}
\end{center}
\end{figure}
\begin{figure}[t]
\centering
\includegraphics[scale=0.27, angle=-90]{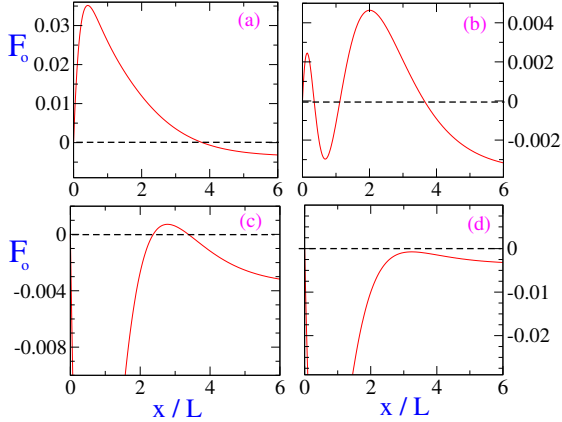}
\begin{center}
\caption{Force vs distance in scaled variables. Here $C_{o} = 0.02$ and $R_{o} = 0.6$ (as in Fig.3.(a)). (a) For  $B_{o} = -0.2$, the only stable solution is that of a bistable spindle. (b) For $B_{o} = -0.4$, both the bistable spindle and the monopolar solution are stable. (c) For $B_{o} = -0.6$, there are two different stable bipolar spindle solutions, while the monopolar solution is also stable. (d) For $B_{o} = -0.8$, only the monopolar solution is stable.}
\label{fig-4}
\end{center}
\end{figure}
\begin{figure}[t]
\centering
\includegraphics[scale=0.27, angle=-90]{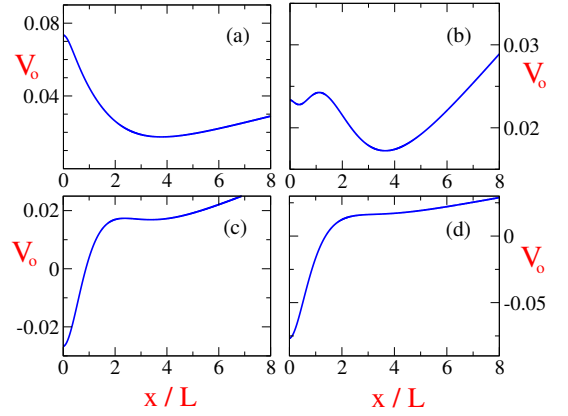}
\begin{center}
\caption{Effective potential energy function ($V_o$) vs scaled distance ($x_o$): Here $C_{o} = 0.02$ and $R_{o} = 0.6$ (Same as Fig.3). (a): $B_{o} = -0.2$, (b) $B_{o} = -0.4$, (c) $B_{o} = -0.6$ and (d) $B_{o} = -0.8$. The corresponding plot of Force vs distance are already shown in Fig.4.} 
\label{fig-5}
\end{center}
\end{figure}

In order to determine the stability boundary we first express Eq.\ref{eq:last} in terms of dimensionless variables. The choice of scaled dimensionless variables are, $x_{o} = x/L$, $F_{o} = \frac{FR_{d}^2}{2NAL^2}$, $R_{o} = R_{d}/L$, $C_{o} = C/A$, $B_{o} = \frac{BR^{2}_{d}}{NAL}$. Then in terms of the these scaled variables, the expression for the dimensionless scaled force is, 

\begin{eqnarray}
F_{o}(x_{o})&=&x_{o}\left[B_{o}e^{-2x_o} + e^{-x_o} - e^{-\sqrt{x_{o}^{2} + R_{o}^{2}}}\right.\nonumber\\
&- &\left.C_{o}\left(\sqrt{x_{o}^{2} + R_{o}^{2}} - x_{o}\right)\right] 
\label{eq:scaledforce-disc}
\end{eqnarray}

This expression is of the form $F(x_{o}) = x_{o} g(x_{o})$ and thus $x_{o} = 0$ is always a solution for Eq.\ref{eq:scaledforce-disc}. From the condition for stability of this particular solution we obtain, 
\begin{equation}
B_{o} = C_{o}R_{o} + e^{-R_1} - 1 \label{eq:pb0}
\end{equation}
For determining the stability boundary for the case when $x_{o}\neq 0$, we set $F_{o}^{'}(x_p) = 0$ and  $F_{o}(x_p) = 0$, which  yields the condition, $g^{'}(x_p) = 0$ and  $g(x_p) = 0$. Using these conditions we obtain,
\begin{eqnarray}
C_{o}&=&\frac{e^{-x_o} + \left(\frac{x_o}{\sqrt{x_{o}^{2} + R_{o}^{2}}} - 2\right) e^{-\sqrt{x_{o}^{2} + R_{o}^{2}}}}{\frac{x_o}{\sqrt{x_{o}^{2} + R_{o}^{2}}} + 2\sqrt{x_{o}^{2} + R_{o}^{2}} - 2x_{o} - 1} \\
B_{o}&=&\frac{ e^{- \sqrt{x_{o}^{2} + R_{o}^{2}}} + C_{o}\sqrt{x_{o}^{2} + R_{o}^{2}} -e^{-x_o} - C_{o}x_{o}   }{e^{-2x_o}}
\end{eqnarray}

These equations are numerically solved to obtain the boundary curves (solid lines) shown in the panels of Fig.3. The stability diagrams in Fig.3 show a variety of scenarios. It exhibits regions where there is bistability with two different stable solutions for bipolar spindle, region where both bipolar spindle and monopolar configuration ($x=0$) is stable, region with just one solution for stable bipolar spindle and a region where no bipolar spindle is stable and only monopolar configuration is stable. The corresponding plots for scaled force as function of centrosome separation is plotted in Fig.4 for these four different regions in the stability diagram. We find that for outward motor-MT forces $(B >0)$, the bipolar spindle is stable as long as $C$ (corresponding to kinetochore-MT interaction strength) is not very large.

By integrating the expression of dimensionless scaled force in Eq. \ref{eq:scaledforce-disc} it is possible to derive an effective dimensionless potential energy function $V_{o}(x_{o})$
\begin{eqnarray}
&V_{o}&=\frac{C_{o}}{3}\left[{\left(\sqrt{x_{o}^{2} + R_{o}^{2}}-x_{o}\right)}^{\frac{3}{2}} - x^{3}_{o}\right] -e^{-x_o}(x_{o} + 1)\nonumber \\
&-&e^{-\sqrt{x_{o}^{2} + R_{o}^{2}}}\left(\sqrt{x_{o}^{2} + R_{o}^{2}} + 1\right)-\frac{B_{o}}{4}e^{-2x_o}(2x_{o} +1)\nonumber
\label{eq:V-disc}
\end{eqnarray}

Fig.5 shows the potential energy landscape for different strengths of $B_o$ for fixed values of other parameters. In particular Fig.5 illustrates that although in the bistable regions two different spindle configurations are stable, in general one of the two corresponds to a global minima in terms of potential energy. The fact that Eq.~(\ref{eq:my-form}) can be derived from an effective potential, $V_o$  is crucial since the functional form of $V_o$ provides insight not only about the stability of the stationary states but also provides the clue about  the dependence on the initial configuration of the system and the long term evolution of the system in presence of noise.  It also rationalizes many of  the observations of the numerical studies in Ref.\cite{raja, ferenz}. First of all it was seen in Ref \cite{raja, ferenz} that the final configuration of the spindle ( on whether it is bipolar or monopolar) is dependent on the initial separation distance. The potential energy landscape provides an explanation on how the final configuration would be dependent on the zones of attraction of the potential landscape with multiple minima. It was also seen in Ref. \cite{raja} that for certain situations, the initial configuration eventually relaxed to a fixed configuration after an intermediate metastable configuration. This can understood in terms of the different depths of the potential minima for the two different configurations. Eventually the system would relax to the global minima in presence of noise in the system. In fact the height of the potential barrier would then determine the time scales of relaxation of the system to its final steady state configuration. 

In order to quantitatively compare the results of the stability diagram obtained here with the numerical studies in Ref.\cite{raja}, we note that they had observed that for parameter values $L= 4$, $N = 46$, $C=10 pN$ and $A = 125 pN$, for $B < 0 pN/\mu m$, monopolar configuration was stable, whereas for $B > 25 pN$, bipolar spindle configuration was stable. In order to compare, first we estimate the typical value of $R_{o} \equiv R_{d}/L$. $R_{o} = \sqrt{N}/L \sim 2$. With this specific choice of $R_1$, we find that bipolar spindle configuration is unstable, while monopolar configuration is stable  for $B=0$ only for  $C_{o} > 0.4$ (Fig.3b), whereas in Ref. \cite{raja} even for $C_{o}\equiv C/A = 0.08$ bipolar spindle was observed to be unstable. We however find that for lower values of B i.e; $( B = -25 pN/\mu m)$, bipolar spindle is unstable like in Ref.\cite{raja}. The quantitative discrepancy between the analytical prediction of this model and the simulation results reported in Ref.\cite{raja} can partially be attributed to our rather crude estimation of the radius of the disc $R_d$. An higher value of $R_d$ would in general result in lower value of $C$ for which stability of bipolar spindle configuration is lost. In general, the parameter range of $A$, $B$ and $C$ along with $L$ and $N$ can be varied over a certain range by altering the physiological conditions within the cell and analogously the scaled variables would span a range of values including $R_d$.

\section{Multiple chromosomes distributed on a ring}  Finally we study the case in which $N$ chromosomes are uniformly distributed on a ring of radius $R_r$. Such a configuration has been shown to be stable in Ref.~\cite{raja}. In such a scenario the expression for the force reads as,
\begin{eqnarray}
F &=&\frac{N x}{2\pi R_{r}}\int_{0}^{2\pi}\frac{R_{r}d\theta}{\sqrt{x^{2}+R_{r}^{2}}} \left( A e^{-\sqrt{x^{2}+R_{r}^{2}}/L} - C \right)\nonumber\\ &+&2Bxe^{-2x/L} 
\label{eq:my-form-r}
\end{eqnarray}
which upon integration yields,
\begin{eqnarray}
F = x\left[\frac{N}{\sqrt{x^{2}+R_{r}^{2}}}\left( A e^{-\sqrt{x^{2}+R_{r}^{2}}/L} - C\right) + 2Be^{-2x/L}\right]
\label{eq:last-r}
\end{eqnarray}
For estimating the radius of the ring $R_r$, we assume that each chromosome occupies $2R_{ch} = 2 $ $\mu m$ length. Therefore $N$ chromosomes distributed on the perimeter of the ring require $R_{r} \sim N/\pi$  $\mu m$.
\begin{figure}[t]
\centering
\includegraphics[scale=0.27, angle=-90]{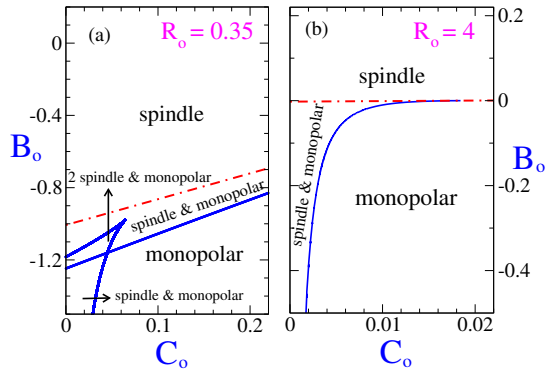}
\begin{center}
\caption{Phase Diagram when chromosomes are distributed on a ring: Here (a) $R_{o} = 0.35$ and (b)  $R_{o} = 4$} 
\label{fig-6}
\end{center}
\end{figure}
The scaled dimensionless variables for this case are, $x_{o} = x/L$, $F_{o} = \frac{F}{NA}$, $R_{o} = R_{r}/L$, $C_{o} = C/A$, $B_{o} = \frac{BL}{NA}$. The corresponding expression for the scaled  force is, 
\begin{eqnarray}
F_{o}(x_{o}) = x_{o} \left[ \frac{e^{-\sqrt{x^{2}+R_{o}^{2}}}}{\sqrt{x^{2}+R_{o}^{2}}}  -\frac{C_o}{\sqrt{x^{2}+R_{o}^{2}}}-2B_{o}e^{-2x_o} \right]
\label{eq:scaledforce2}
\end{eqnarray}
For the solution $x_{o} = 0$, the condition for stability yields, 
\begin{equation}
B_{o} = \frac{C_o - e^{-R_1} }{2R_1}  \label{eq:pb0r}
\end{equation}
 that is represented in the panels of Fig.~6 as a dashed-dotted line.
For determining the phase boundary for the case when $x_{o}\neq 0$, we again set $\frac{d F}{dx}\big|_{x_{o}}  = 0$ and  $F_{o}(x_{o}) = 0$, to obtain,
\begin{eqnarray}
C_{o}&=& \frac{e^{-\sqrt{x_{o}^{2} + R_{o}^{2}}}\left[2 - \frac{x_o}{x_{o}^{2}+R_{o}^{2}} -\frac{x_o}{\sqrt{x_{o}^{2} + R_{o}^{2}}}\right]}{2 - \frac{x_o}{x^{2}_{o} + R_{o}^{2}}}\\
B_{o}&=&\frac{- e^{2x_o}}{2\sqrt{x_{o}^{2} + R_{o}^{2}}}\left[e^{-\sqrt{x_{o}^{2}+ R_{o}^{2}}} - C_{o}\right]
\end{eqnarray}
These equations are numerically solved to obtain the boundary curves (solid lines) in Fig.~6 separating the stable and unstable region. In particular, Fig.~6.(a) exhibits regions of bistability for which two different solutions for bipolar spindle are stable and regions where both bipolar spindle and monopolar configuration are stable. In Fig.~6(b) we choose a value of $R_{r} = 4$, obtained by estimating $N= 46$ and $L = 4$ as done in Ref.\cite{raja}. We find that for this ring geometry, even for very low value of $C$, the bipolar spindle becomes unstable as was observed in Ref.\cite{raja}.
As in the previous case, by integrating the expression of dimensionless scaled force in Eq. \ref{eq:scaledforce2}, it possible to associate an effective potential energy function $V_{s}(x_{0})$, as a function  of centrosome separation distance in the scaled variable $x_o$. This expression reads 
\begin{eqnarray}
V_{o}=C_{o}\sqrt{x_{o}^{2} + R_{o}^{2}} + \frac{B_{o}}{2}e^{-2x_o}(2x_{o} + 1) + e^{-\sqrt{x_{o}^{2} + R_{o}^{2}} }
\label{eq:V-ring}
\end{eqnarray}

\section{Summary and discussion} In summary, we have studied and analyzed the mechanical stability of mitotic spindle structure within the framework of a minimal model proposed in Ref. \cite{raja, ferenz} which incorporates the interactions between Chromosomes with MTs nucleated from the centrosome, the sliding forces generated by motor proteins on overlapping MT filaments, and the interactions of the MTs with kinetochore machinery within the cell. Having assumed that the chromosomes are {\it a priori} distributed homogeneously on a disc or a ring in the mid-plane between the two centrosomes, we obtain a closed form analytic expression for the forces acting between pair of centrosomes. The assumption of such an arrangement of chromosomes is consistent with earlier numerical and experimental studies \cite{raja,cell,toroid}.  We analyze the stability of the spindle configurations  and obtain the corresponding stability diagram by invoking the condition of {\it stable} mechanical equilibrium apart from the balance of forces. This stability diagram is expressed in terms of  dimensionless parameters which are essentially ratios of the strength of the different interaction forces. This stability diagram allows us to quantify the impact of the different interactions at play in determining the final configuration of the spindle. We find that in general both for the case of chromosomes arranged on a ring and disc, if there is a net outward sliding forces by motors $(B >0)$ then  bipolar spindle  is stable, whereas for sufficiently high inward sliding forces of motor $(B<0)$, monopolar configuration is stable. We also found that for sufficiently high value of $C$, even for outward motor sliding force $( B >0)$, the bipolar spindle configuration can be unstable. Interestingly, when $B <0$, below a threshold value of $C$ corresponding to kinetochore-MT interaction forces, there are regimes of bistability  characterized by coexistence of two different spindle length, and also stability of the bipolar spindle configuration and monopolar configuration. The origin of this bistable behavior can be associated with multiple minima in the effective potential energy function that can be derived from the expression of total interaction force. In fact it also allows us to rationalize the observation in Ref. \cite{raja,ferenz}, where it was seen that the final stable configuration was seen to be dependent on the initial separation distance.  It would be interesting to investigate whether the prolonged lifetime in the bipolar spindle configuration to the eventual {\it monopolar} state observed in Ref. \cite{raja} can be attributed to transition from shallow potential minima corresponding to spindle configuration to deeper potential energy minima associated with the monopolar configuration. 

\section{Acknowledgments}
SM acknowledges DBT RGYI Project No. BT/PR6715/GBD/27/463/2012 for financial support, and S. Dietrich (MPI, Stuttgart) for travel grant and stay during research visit to MPI, Stuttgart.


\begin{thebibliography}{}

\bibitem{cell} B. Alberts at. al , {\sl Molecular Biology of the cell} ( Garland Science, New York, 2002, 4th ed)

\bibitem{howard} J. Howard, {\sl Mechanics of Motor Proteins and the Cytoskeleton} (Sinauer, Sunderland, 2001).

\bibitem{raja} S. Sutradhar, S. Basu, and R. Paul, Phys. Rev. E. {\bf 92}, 042714 (2015).

\bibitem{ferenz} N. P. Ferenz, R. Paul, C. Fagerstrom, A. Mogilner, and P. Wadsworth, Curr. Biol. {\bf 19}, 1833 (2009).

\bibitem{A1} C.L. Reider, E. A. Davison, L. C. Jensen, L. Cassimeris, and E. D. Salmon, J. Cell. Biol. {\bf 103}, 581 (1986).

\bibitem{A2} K. Ke, J. Cheng, and A. J. Hunt, Curr. Biol. {\bf 19}, 807 (2009).

\bibitem{C1} M. Kirschner and T. Mitchison, Cell. {\bf 45}, 329 (1986).

\bibitem{C2} S. L. Kline-Smith, S. Sandall, and A. Desai, Curr. Opin. Cell. Biol. {\bf 17}, 35 (2005).

\bibitem{C3} I. M. Cheeseman and  A. Desai, Nat. Rev. Mol. Cell. Biol. {\bf 9}, 33 (2008).

\bibitem{B1} L. A. Cameron, G. Yang, D. Cimini, J. C. Canman, O. Kisurina-Evgenieva, A. Khodjakov, G. Danuser, and E. D. Salmon, J. Cell. Biol. {\bf 173}, 173 (2006).

\bibitem{B2} M. E. Tanenbaum, L. Macurek, N. Galjart, and R. H. Medema, EMBO J. {\bf 27}, 3235 (2008).

\bibitem{B3} S. Muhuri, I. Pagonabarraga, J. Casademunt,  EPL. {\bf 98}, 68005 (2012).

\bibitem{astral-mt} J. L. Carminati and T. Stearns, J. Cell. Biol. {\bf 138}, 629 (1997).

\bibitem{exp-length} M. Dogterom and S. Leibler,  Phys. Rev. Lett. {\bf 70}, 1347 (1993).

\bibitem{f-chromo1} C. L. Rieder and E. D. Salmon, J. Cell. Biol. {\bf 124}, 223 (1994).

\bibitem{f-chromo2} A. P. Joglekar, D. C. Bouck, J. N. Molk, K. S. Bloom, and E. D. Salmon, Nat. Cell Biol. {\bf 8}, 581 (2006).

\bibitem{raja2} S. Sau, S. Sutradhar, R. Paul, and P. Sinha, PLoS One. {\bf 9}, e101294 (2014).

\bibitem{toroid} V. Magidson, C. B. O' Conell, J. Loncarek, R. Paul, A. Mogilner, and A. Khodjakov, J. Cell. Sci. {\bf 146}, 555 (2011).


\end{thebibliography}
\end{document}